\documentclass[aps,showpacs,superscriptaddress,twocolumn]{revtex4}
\usepackage[intlimits]{amsmath}
\usepackage{amssymb}
\usepackage{graphicx}
\usepackage{booktabs}
\usepackage{mathrsfs,slashed}
\usepackage{hyperref}
\usepackage{epsfig}
\usepackage{color}
\newcommand{\bea}{\begin{eqnarray}}
\newcommand{\eea}{\end{eqnarray}}

\def\gsim{\stackrel{\scriptstyle >}{\phantom{}_{\sim}}}

\begin{document}
%\title{Third family of compact stars in a relativistic mean-field model with phase transition}
%\title{Third family of massive compact stars for a deconfinement transition with a mixed phase construction} 
%\title{How robust is the existence of mass-twin stars against pasta phase effects?} 
%\title{How robust is a third family of compact  stars against pasta phase effects?} 
\title{Robustness of third family solutions for hybrid stars against mixed phase effects} 

\author{A. Ayriyan}
\email{ayriyan@jinr.ru}
\affiliation{Laboratory for Information Technologies,
	Joint Institute for Nuclear Research,
	Joliot-Curie street 6,
	141980 Dubna, Russia}
\author{N.-U.~Bastian}
\email{niels-uwe.bastian@ift.uni.wroc.pl}
\affiliation{Institute of Theoretical Physics, 
	University of Wroclaw, 
	Max Born place 9, 
	50-204 Wroclaw, Poland}
\author{D.~Blaschke}
\email{blaschke@ift.uni.wroc.pl}
\affiliation{Institute of Theoretical Physics, 
	University of Wroclaw, 
	Max Born place 9, 
	50-204 Wroclaw, Poland}
\affiliation{Bogoliubov Laboratory for Theoretical Physics,
	Joint Institute for Nuclear Research,
	Joliot-Curie street 6,
	141980 Dubna, Russia}
\affiliation{National Research Nuclear University (MEPhI),
	Kashirskoe Shosse 31,
	115409 Moscow, Russia}
\author{H. Grigorian}
\email{hovikgrigorian@gmail.com}
\affiliation{Laboratory for Information Technologies,
	Joint Institute for Nuclear Research,
	Joliot-Curie street 6,
	141980 Dubna, Russia}
\author{K. Maslov}
\email{maslov@theor.mephi.ru}
\affiliation{Bogoliubov Laboratory for Theoretical Physics,
	Joint Institute for Nuclear Research,
	Joliot-Curie street 6,
	141980 Dubna, Russia}
\affiliation{National Research Nuclear University (MEPhI),
	Kashirskoe Shosse 31,
	115409 Moscow, Russia}
\author{D. N. Voskresensky}
\email{d.voskresensky@gsi.de}
\affiliation{Bogoliubov Laboratory for Theoretical Physics,
	Joint Institute for Nuclear Research,
	Joliot-Curie street 6,
	141980 Dubna, Russia}
\affiliation{National Research Nuclear University (MEPhI),
	Kashirskoe Shosse 31,
	115409 Moscow, Russia}

%%%%%%%%%%%%%%%%%%%%%%%%%%%%%%%%%%%
\date{\today}
\begin{abstract}
We investigate the robustness of third family solutions for hybrid compact stars with a quark matter core that correspond to the occurrence of high-mass twin stars against a softening of the phase transition by means of a construction that mimics the effects of pasta structures in the mixed phase.
We consider a class of hybrid equations of state that exploits a relativistic mean-field model for the hadronic as well as for the quark matter phase. 
We present parametrizations that correspond to branches of high-mass twin star pairs with maximum masses between $2.05~M_\odot$  and $1.48~M_\odot$ having radius differences between 3.2 and 1.5 km, respectively.
When compared to a Maxwell construction with a fixed value of critical pressure $P_c$, the effect of the mixed phase construction consists in the occurrence of a region of pressures around $P_c$ belonging to the coexistence of hadronic and quark matter phases between the onset pressure at $P_H$ and the end of the transition at $P_Q$.
The maximum broadening which would still allow mass twin compact stars is found to be 
$(P_Q-P_H)_{\rm max} \approx P_c$ for all parametrizations within the present class of models. 
At least the heavier of the neutron stars of the binary merger GW170817 could have been a member of the  third family of hybrid stars. 
We present the example of another class of hybrid star equations of state for which the appearance of the third family branch is not as robust against mixed phase effects as that of the present work.
\end{abstract}
%\begin{keyword}
%quantum chromodynamics, chiral symmetry, finite density/temperature, 
%diquarks, nucleons
%\end{keyword}
\maketitle
%\tableofcontents

%%%%%%%%%%%%%%%%%%%%%%%%%%%%%%%%%%%%%%%%%%%%%%%%%
%%%%%%%%%%%%%%%%%%%%%%%%%%%%%%%%%%%%%%%%%%%%%%%%%

\section{Introduction}

Recently, significant attention has been devoted to the qualitative features of the mass-radius diagram for compact stars and to the conclusions that could be drawn from the diagram regarding the properties of the equation of state (EoS) of degenerate matter under neutron star (NS) conditions.
This attention has been driven by the observation of high masses $\sim 2~M_\odot$ of pulsars at high accuracy \cite{Demorest:2010bx,Fonseca:2016tux,Antoniadis:2013pzd}, as well as constraints on the neutron star radii.
A new level of quality will be achieved when results for simultaneous mass and radius measurements from  the NICER experiment \cite{NICER1} will become available and when more gravitational wave signals from NS-NS mergers like GW170817 \cite{TheLIGOScientific:2017qsa} will be detected that eventually will show a ringdown signal and not just point to a prompt collapse.

One of the challenging questions in this context is the possibility of a phase transition to exotic forms of matter, in particular to deconfined quark matter in the interior of massive compact stars. 
As was analyzed in Ref.~\cite{Alford:2013aca}, the phenomenology of hybrid star sequences in the mass-radius ($M-R$) diagram of compact stars may be classified into four groups in the so-called "phase diagram of hybrid stars" which is spanned by the critical pressure ($P_c$) and the jump in the energy density ($\Delta \varepsilon$) characterizing the deconfinement phase transition. 
Of particular interest is the case of stable hybrid star branches that are disconnected from the normal NS branch and thus form a ``third family'' of compact stars \cite{Gerlach:1968zz} that show a strong phase transition with a large $\Delta \varepsilon \gsim 0.5 \varepsilon_c$ \cite{Seidov:1971,Schaeffer:1983}, where $\varepsilon_c$ is the energy density at the onset of the deconfinement phase transition. 
As has been shown in \cite{Blaschke:2013ana} for several realistic hybrid star EoS models, this case can occur also for pulsars with high masses where then the phenomenon of ``high-mass twin'' (HMT) stars is possible \cite{Benic:2014jia}. 
The HMT phenomenon is not quite as exotic as it may seem at first glance. It can also be obtained within the multi-polytrope scheme for the high-density EoS \cite{Alvarez-Castillo:2017qki} and meets the constraints analyzed by Hebeler et al.~\cite{Hebeler:2013nza}. 
 
The question we would like to answer in this work concerns the robustness of the HMT phenomenon when the deconfinement phase transition in the compact star interior is not described by a Maxwell construction with a jump in energy density at constant pressure for which the speed of sound in matter vanishes, but when the transition proceeds rather with a finite pressure gradient. 
In the latter case, a mixed phase can be realized in the hybrid star structure for which the speed of sound develops a dip but does not vanish. Such a case may occur when the phase transition proceeds under global charge conservation for systems with several conserved charges \cite{Glendenning:1992vb}. 
A mixed phase appears provided the surface tension between quark and hadron phases does not exceed a critical value estimated as $\sigma_c\sim 60~\mathrm{MeV fm^{-2}}$; see 
\cite{Heiselberg:1992dx,Voskresensky:2002hu}.
It can lead to the formation of structures and such a mixed phase is then dubbed "pasta phase". 
See, e.g., Ref.~\cite{Yasutake:2014oxa} for more details and a recent calculation with modern hadronic and quark matter EoSs.

\section{Hybrid star equation of state}

\subsection{Hadronic phase}
\label{ssec:hp}

For description of the hadronic phase we use a relativistic mean-field (RMF) model, the phenomenological KVORcut02 EoS \cite{Maslov:2015wba}. 
It was constructed with hadron masses and coupling constants dependent on the scalar field. 
In this model all the hadron masses are decreasing functions of the mean value of the scalar field in medium with a common rate in accordance with the idea of partial restoration of the chiral symmetry and the Brown-Rho scaling conjecture.
The effective coupling constants enter the expressions for the EoS only in ratios with effective masses, and the dependence of these ratios on the scalar mean field is chosen to describe experimental constraints the best way. 
Here we disregard the possibility of a finite hyperon abundance following the argument given in Ref.~\cite{Maruyama:2007ss} that hyperons can be completely suppressed in the mixed phase because the hadronic phase is positively charged.

On the basis of the KVOR model \cite{Kolomeitsev:2004ff} a family of models (labeled KVORcut) was constructed in \cite{Maslov:2015wba}, implementing a version of the stiffening mechanism developed in \cite{Maslov:cut}. 
This mechanism allows one to make a given EoS stiffer at suprasaturation densities without altering it at lower densities, so the saturation properties remain unchanged. KVORcut02 is the stiffest of the EoS of this family, predicting a maximum NS mass of $2.26~M_\odot$ and a maximum radius of $R \simeq 13.9$ km.

\subsection{Quark matter phase}
\label{ssec:qp}

For the description of the quark matter phase we choose a recently developed relativistic density functional approach \cite{Kaltenborn:2017hus}. 
It is capable of modeling effects of confinement by diverging quark masses at low densities and includes an effective screening of the string tension $\sigma_0$ in the confining interaction
\begin{eqnarray}
\sigma(n)=\sigma_0~\Phi(n)~.
\end{eqnarray}
The density-dependent screening is described with the same functional as the excluded volume effect in a recent version \cite{Typel:2016abc} of the density-dependent relativistic mean-field model DD2 \cite{Typel:2009sy}
\begin{eqnarray}
\Phi(n)=\exp\left[-\alpha\left(n/{\rm fm}^{-3}\right)^2\right]~.
\end{eqnarray} 
In place of the excluded volume parameter of that model, the parameter $\alpha$ appears in the string-flip model.

A repulsive vector interaction is included with a higher-order density dependence so that a relatively soft behavior at the deconfinement transition can be reconciled with a sufficiently fast stiffening of the EoS in order to sustain the $2M_\odot$ constraint for hybrid stars with a quark matter core.
The former is necessary to obtain a large enough density jump at the transition which induces a gravitational instability as a condition for the appearance of a separated branch of hybrid stars in the mass-radius diagram that would form a ``third family'' of compact stars.
The higher order density-dependent repulsion term requires a density-dependent correction factor in order to maintain causality of the EoS so that for the speed of sound it holds that $c_s^2 \le 1$.
For details, see Ref.~ \cite{Kaltenborn:2017hus}.

\subsection{Phase transition construction}
\label{ssec:ptc}

When the hadronic and quark matter phases are described with EoSs given by relations between the pressure and chemical potential (for $T=0$, which is relevant for the NS modeling) $P_{H}(\mu)$ and $P_{Q}(\mu)$ respectively, one can find the critical value of the baryochemical potential $\mu_c$ from the condition of equal pressures, 
\begin{equation}
P_{Q}\left(\mu_{c}\right)=P_{H}\left(\mu_{c}\right) = P_c,
\end{equation}
for which the phases are in mechanical equilibrium with each other. 
The value $P_c$ defines the Maxwell construction. 
Quantities characterizing the quark-, hadron-, and mixed phases are denoted by the subscripts $Q$, $H$, and $M$, respectively.

Assuming the surface tension to be smaller than the critical value $\sigma_c$, a mixed phase could have an influence on the compact stars structure.
The adequate description of the physics of pasta phases is a complicated problem which requires one to take into account sizes and shapes of structures as well as transitions between them.
It has been dealt with in the literature within different methods and approximations \cite{Voskresensky:2002hu,Maruyama:2007ss,Yasutake:2014oxa,Watanabe:2003xu,Horowitz:2005zb,Horowitz:2014xca,Newton:2009zz}.

\begin{figure}[!htb]
	\includegraphics[width=1.1\columnwidth]{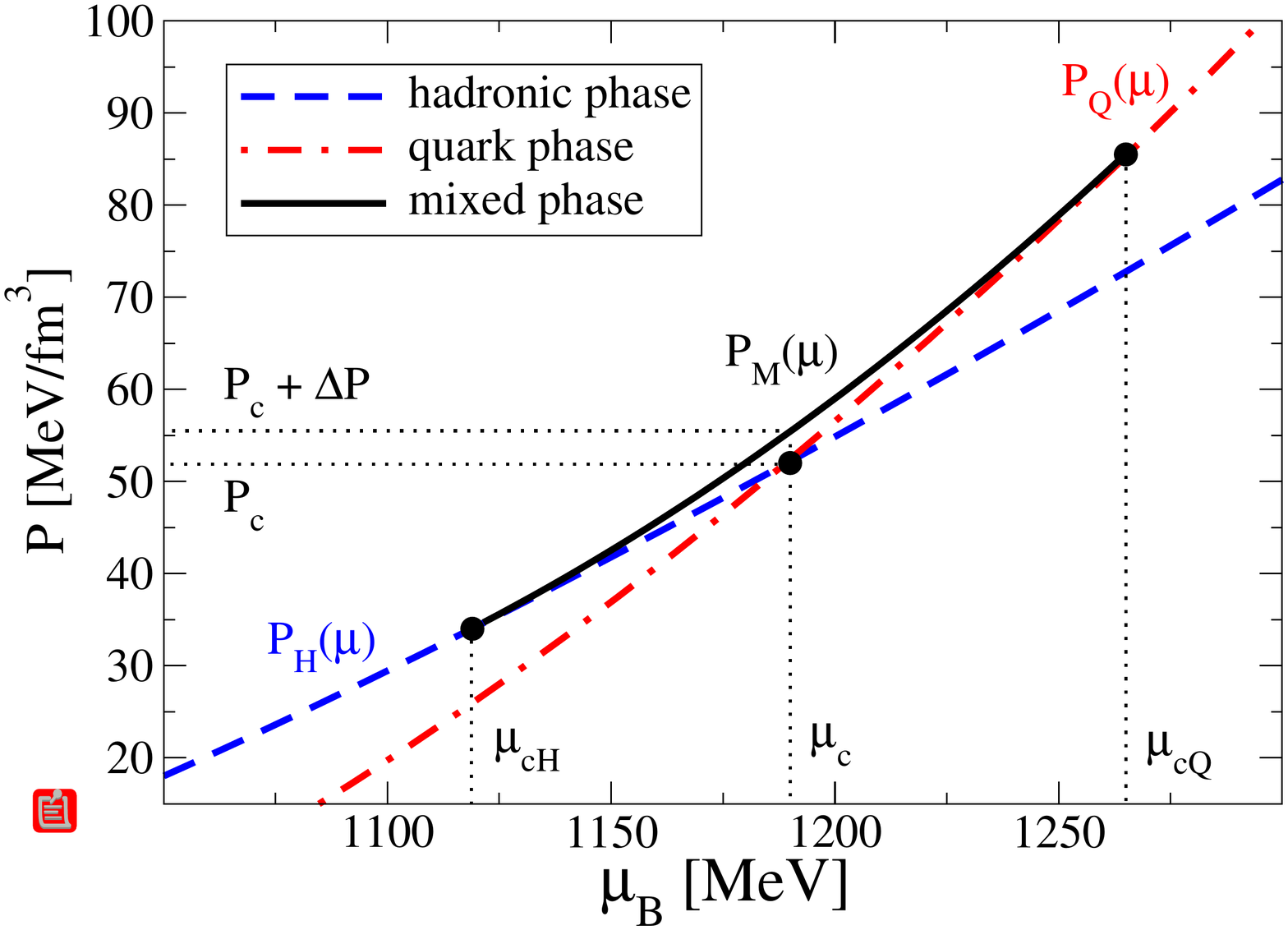}
	\caption{The mixed phase construction (solid line) for given hadronic matter (dashed line) and quark 	matter (dash-dotted line)  EoS models employed in this work.
	The Maxwell construction corresponds to switching at $\mu_c$ from the hadronic EoS for 
	$\mu_B<\mu_c$ to the quark matter EoS for $\mu_B>\mu_c$.}
	\label{fig1}       % Give a unique label
\end{figure}

Here, instead of the full solution we would like to use a simple modification of the Maxwell construction which mimics the result of pasta matter studies \cite{Ayriyan:2017tvl}.  
This means that the pressure as a function of the baryon density in the mixed phase is not constant but rather a monotonically rising function.
We are assuming that close to the phase transition point (that would be obtained using the Maxwell construction) the EoS of both phases are changing due to finite size and Coulomb effects (see Fig.~\ref{fig1}), so that the effective mixed phase EoS $P_{M}(\mu)$ could be described in the parabolic form 
\begin{equation}
P_{M}(\mu)=a (\mu-\mu_{c})^2+ b (\mu-\mu_{c})+P_c+\Delta P.
\label{p_mu}
\end{equation}
Here we have introduced the pressure shift $\Delta P$ at $\mu_{c}$ as a free parameter of the model which determines the mixed phase pressure at this point
\begin{equation}
P_{M}\left(\mu_{c}\right)=P_c+\Delta P = P_M.
\end{equation}
The parameter $\Delta P$ shall be related to the largely unknown surface tension between the hadronic and quark matter phases of the strongly interacting system at the phase transition. Infinite surface tension corresponds to the Maxwell construction and thus to $\Delta P=0$, while a vanishing surface tension is the case of a Gibbs construction under global charge conservation (also called Glendenning construction \cite{Glendenning:1992vb}) that for known examples looks similar to the results of our mixed phase construction for the largest values considered here, $\Delta_P\sim 0.07 -- 0.10$.
A more quantitative relation between $\Delta P$ and the surface tension would require a fit of the mixed phase parameter to a pasta phase calculation for given surface tensions. Such a calculation is under way along the lines of Ref.~\cite{Yasutake:2014oxa}, but goes beyond the scope of the present work. 

According to the mixed phase construction shown in Fig.~\ref{fig1} we have two critical chemical potentials: $\mu_{cH}$ for the transition from $H$~phase to $M$~phase and $\mu_{cQ}$ for the transition from $M$~phase to $Q$~phase. 
Together with  the coefficients $a$ and $b$ from Eq.~\eqref{p_mu} these are four unknowns which 
shall be determined from the four equations for the continuity of the pressure 
\begin{eqnarray}
\label{PH}
	P_{M}(\mu_{cH}) & = & P_{H}(\mu_{cH}) = P_H,\\
	P_{M}(\mu_{cQ}) & = & P_{Q}(\mu_{cQ}) = P_Q,
\label{PQ}
\end{eqnarray}
and of the baryon number density $n(\mu)=dP(\mu)/d\mu$,
\begin{eqnarray}
\label{nH}
	n_{M}(\mu_{cH}) & = & n_{H}(\mu_{cH}),\\
	n_{M}(\mu_{cQ}) & = & n_{Q}(\mu_{cQ}).
\label{nQ}
\end{eqnarray}

From Eqs.~ (\ref{PH}) and (\ref{PQ}) for the pressure, $a$ and $b$ are found as
\begin{eqnarray}
	a & = & \frac{1}{\mu_{cQ}-\mu_{cH}}\left(\frac{P_{Q}-P_{M}}{\mu_{cQ}-\mu_{c}}
	- \frac{P_{H}-P_{M}}{\mu_{cH}-\mu_{c}}\right)\,,\\
	b & = & \frac{P_{M}-P_{H}}{\mu_{c}-\mu_{cH}}
	+\frac{P_{Q}-P_{M}}{\mu_{cQ}-\mu_{c}}
	-\frac{P_{Q}-P_{H}}{\mu_{cQ}-\mu_{cH}}~,
\end{eqnarray}
and can be eliminated from the set of equations for the densities.
By solving the remaining Eqs.~ (\ref{nH}) and (\ref{nQ}) for the densities numerically one can find the values for the critical chemical potentials $\mu_{cH}$ and $\mu_{cQ}$.

The procedure for ``mimicking pasta phases'' described in this subsection is general and can be superimposed on any hybrid EoS which was obtained using the Maxwell construction of a first-order phase transition from a low-density to a high-density phase with their given EoSs.
A certain limitation occurs in the case of sequential phase transitions (see, e.g., Refs.~\cite{Blaschke:2008br}  and \cite{Alford:2017qgh}) when the broadening of the phase transition region as shown in Fig.~\ref{fig1} would affect the adjacent  transition. 
This case shall be discussed elsewhere \cite{Ayriyan:2017}.

%%%%%%%%%%%%%%%%%%%%%%%%%%%%%%%%%%%%%%%%%%%%%%%%%%%%% 

\section{Results and discussion}
\label{results}

In this section we present and discuss results obtained for the case of the quark deconfinement transition when the hadronic EoS is given by the relativistic mean-field model described in Sec.~\ref{ssec:hp} and the quark matter EoS is chosen by the relativistic density-functional approach motivated by the string flip model as characterized in Ssec.~\ref{ssec:qp}.
First we show the effect of broadening the phase transition region that the introduction of the free parameter $\Delta P$ of the mixed phase construction has on the hybrid star EoS.
This is done for selected parameter values of the quark matter model.
Subsequently we investigate the corresponding mass-radius relations for compact stars.
We are particularly interested in the presence of a third family branch of hybrid stars with a large quark matter core, disconnected from the usual hadronic branch of compact stars to which eventually a short hybrid star branch with small mixed phase cores might be connected.
The possibility of such solutions may become of particular interest for the phenomenology of gravitational wave signals from NS-NS merger events. 
They would allow for metastable high-mass hybrid star solutions as a particular case in delayed collapse scenarios of the merger evolution.

Furthermore, a new aspect of this work is to quantify a critical broadening effect for which the third family branch of hybrid star solutions would get connected to the main neutron star branch.
That is, we will identify the maximum for the numerical value of the dimensionless broadening parameter $\Delta_P = \Delta P/P_c$ that would still allow for the appearance of a third family branch and therefore high-mass twin star solutions.

%%%%%%%%%%%%%%%%%%%%%%%%%%%%%%%%%%%%%%%%%%%%%%%%%%%%
\subsection{Hybrid star EoS with mixed phase construction}

\begin{figure}[!htb]
	\includegraphics[width=1.1\columnwidth]{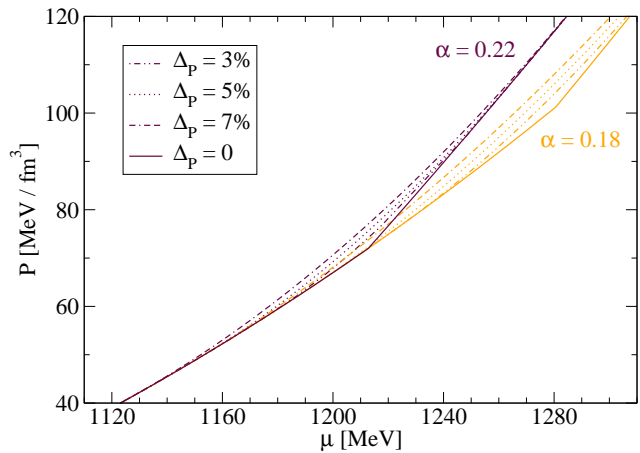}
	\caption{%
		Hybrid EoS according to the mixed phase construction of Fig.~\ref{fig1} described in the text, for the string screening parameter values $\alpha=0.18$ and 0.22. 
		For the mixed phase parameter $\Delta_P=\Delta P/P_C$ the values 0.0 (Maxwell construction limit) 0.03, 0.05, and 0.07 are chosen.} 
	\label{fig:eos}
\end{figure}

In Fig.~\ref{fig:eos} we present results of the mixed phase construction outlined in Sec.~\ref{ssec:ptc} where inputs are the fixed hadronic EoS as described in Sec.~\ref{ssec:hp} and the quark matter EoS of the string-flip model characterized in Subsect.~\ref{ssec:qp}. 
For the latter, we vary the value of the excluded volume parameter $\alpha$, which corresponds to an effective reduction of the string tension in a dense medium. From the parameter range explored in \cite{Kaltenborn:2017hus} we focus on the values $\alpha=0.18$, $0.20$, $0.22$, $0.24$, and $0.30$.
We show results for pressure vs. chemical potential in Fig.~\ref{fig:eos}.

%%%%%%%%%%%%%%%%%%%%%%%%%%%%%%%%%%%%%%%%%%%%%%%%%%%
With the choice $\alpha=0.18$ $(0.22)$, we obtain the phase transition by  Maxwell construction ($\Delta_P=0$) at $P_c=101$ MeV/fm$^3$ (71 MeV/fm$^3$). 
The moderate choice of $\Delta_P\sim 5 \%$ leads to a lowering of the onset pressure for the phase transition by $\sim 40 \%$.  
The corresponding region of pressures around $P_c$ belonging to the coexistence of hadronic and quark matter phases between the onset pressure at $P_H$ and the end of the transition at $P_Q$ is found to be $P_Q-P_H \approx P_c$ for all parametrizations within the present class of models. 
%%%%%%%%%%%%%%%%%%%%%%%%%%%%%%%%%%%%%%%%%%%%%%%%%%%%

\begin{figure}[!thb]=
	\includegraphics[width=1.1\columnwidth]{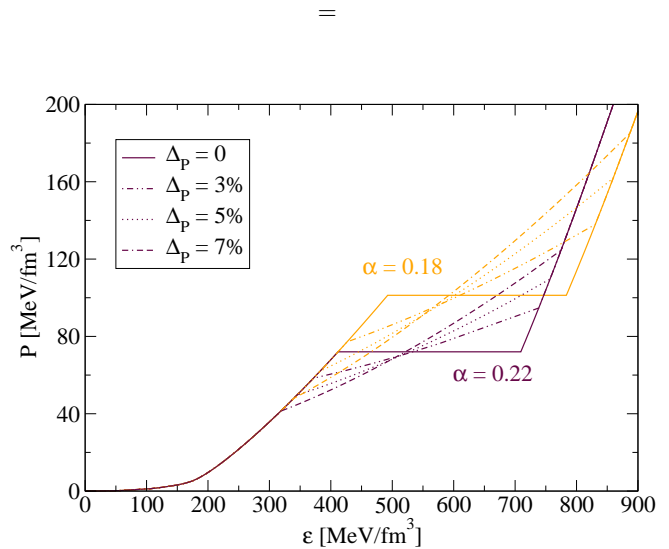}
	\caption{Same as Fig.~\ref{fig:eos} for pressure vs. energy density.} 
	\label{fig:p-eps}
\end{figure}

%%%%%%%%%%%%%%%%%%%%%%%%%%%%%%%%%%%%%%%%%%%%%%%%%%%%

\begin{figure}[!htb]
	\includegraphics[width=1.1\columnwidth]{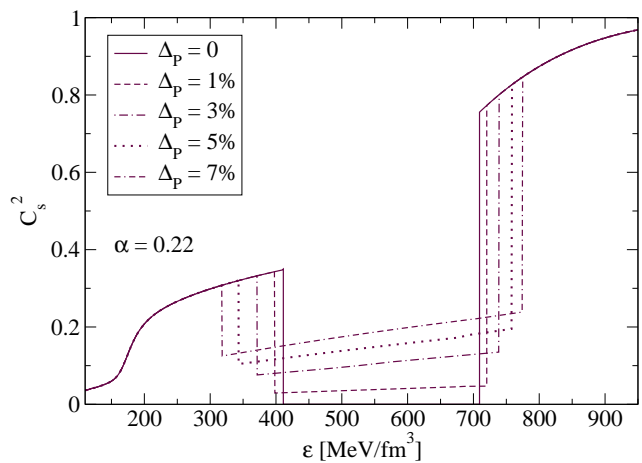}
	\caption{%
	Speed of sound squared as a function of energy density, derived from the set of EoS in Fig.~\ref{fig:p-eps} for $\alpha=0.22$ and different mixed phase parameters $\Delta_P=0.0--0.07$.} 
	\label{fig:cs2}
\end{figure}

In Fig.~\ref{fig:p-eps} we show the pressure vs. energy density corresponding the results shown in Fig.~\ref{fig:eos}.
Note that the jump in energy densities for the Maxwell construction amounts to about $70\%--75\%$ of the critical energy density at the phase transition onset. 
An interesting question concerns the relation between the parameter $\Delta_P$ and the surface tension between quark and hadron phases.  
To extract a more quantitative relationship between the surface tension and the parameter $\Delta_P$ requires a calculation of the pasta structures in the mixed phase region for the given hadron and quark EoSs, including the effects of the Coulomb energy for given surface tensions, then fitting by a suitable value of the $\Delta_P$ parameter of the mixed phase model.
Such a calculation is beyond the scope of the present work and shall be reported elsewhere. The best fit may require a nonparabolic interpolation function as used in, e.g., Refs. \cite{Alvarez-Castillo:2014dva,Alvarez-Castillo:2017xvu}.
The effects of the mixed phase construction are a broadening of the coexistence region of the phases and a nonzero speed of sound in it, as shown in Fig.~\ref{fig:cs2}.

%%%%%%%%%%%%%%%%%%%%%%%%%%%%%%%%%%%%%%%%%%%%%%%
\subsection{Mass-radius diagram}
\begin{figure}[!htb]
	\includegraphics[width=1.1\columnwidth]{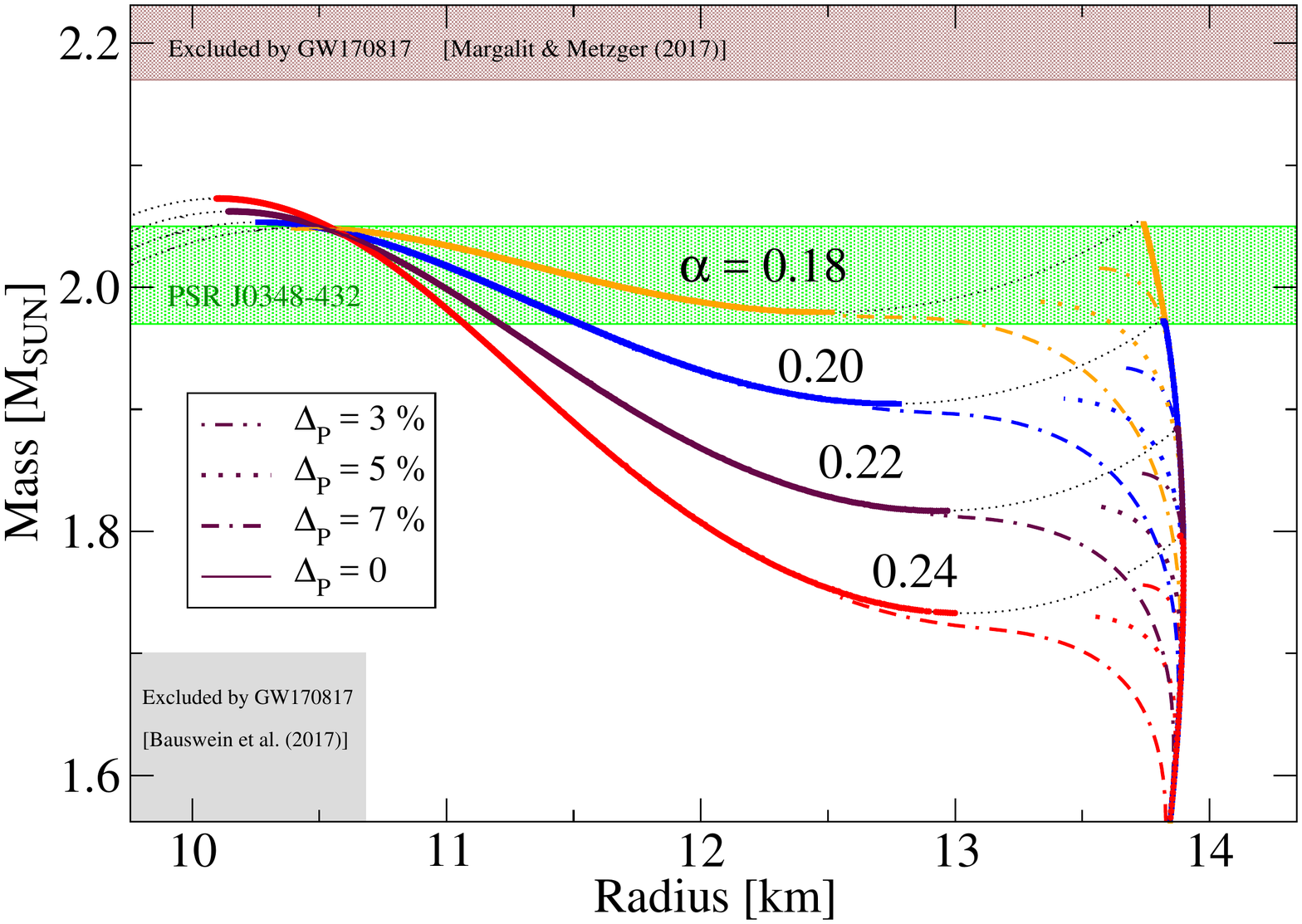}
	\caption{Mass-radius diagram for the hybrid star sequences of this work for different values of the screening parameter $\alpha=0.18$, 0.20, 0.22, 0.24 and varying value of the mixed phase parameter $\Delta_P=0$ (solid lines; the disconnected NS and third family branches are joined by a thin dotted line representing unstable configurations to guide the eye), 0.03 (dash-double-dotted lines), 0.05 (dotted lines) and 0.07 (double-dash-dotted lines). The critical value for the transition from the third family case to a connected hybrid star branch is $\Delta_P\sim 0.06$ in all cases.}
	\label{fig:M-R}
\end{figure}

%%%%%%%%%%%%%%%%%%%%%%%%%%%%%%%%%%%%%%%%%%%%%%%
The EoS shown in Figs.~\ref{fig:eos} - \ref{fig:cs2} has been used to solve the Tolman-Oppenheimer-Volkoff (TOV) equations for structure and stability of compact star configurations. 
The results for the mass-radius relationships (sequences) corresponding to the present class of EoS models are shown in Fig.~\ref{fig:M-R}.
Increasing the $\alpha$ parameter from $0.18$ to $0.24$ leads to a lowering of the onset mass of the phase transition from $2.05~M_\odot$ to $1.80~M_\odot$ for the Maxwell construction case. 
Dotted lines correspond to unstable configurations which shall not be realized in nature and are shown only to guide the eye.
The occurrence of such a gap in the mass-radius sequence defines the existence of a third family of compact stars, here they are hybrid stars with a quark matter core.
The effect of the mixed phase construction is to lower the onset mass of the transition and to allow for a short sequence of hybrid stars with a small mixed phase core which is connected to the second family branch of ordinary neutron stars.
In the classification scheme of Ref.~\cite{Alford:2013aca} this would correspond to the case ``B'' for both, disconnected and connected hybrid star branches for the same EoS.
The critical value of pressure shift which would join these two branches and thus eliminate the third family case is found to be $\Delta_P\sim 6\%$; see Fig.~\ref{fig:M-R}.   

%%%%%%%%%%%%%%%%%%%%%%%%%%%%%%%%%%%%%%%%%%%%%%%%%%%%
In Fig.~\ref{fig:M-R} we show together with the $M-R$ sequences also the $2~M_\odot$ mass constraint of Ref.~\cite{Antoniadis:2013pzd} and two constraints \cite{Bauswein:2017vtn,Margalit:2017dij} which have recently been extracted from  GW170817 under the assumption of no direct collapse in this binary neutron star merger event. 
%%%%%%%%%%%%%%%%%%%%%%%%%%%%%%%%%%%%%%%%%%%%%%%

\begin{figure}[!htb]
	\includegraphics[width=1.1\columnwidth]{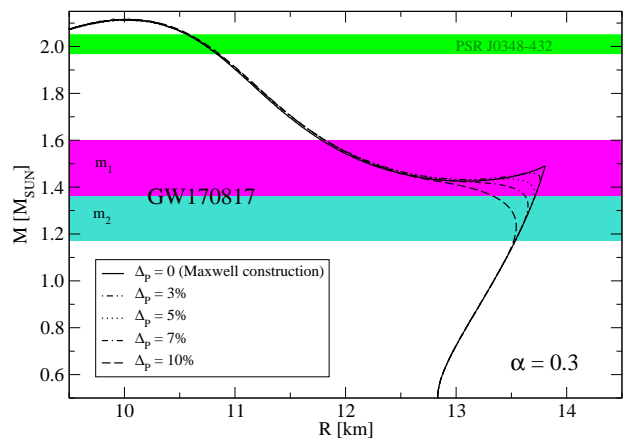}
	\caption{Mass-radius diagram for the hybrid star sequences of this work for the screening parameter $\alpha=0.3$ and varying value of the mixed phase parameter: $\Delta_P=0$ (solid line), 0.03 (dash-double-dotted line), 0.05 (dotted line), 0.07 (double-dash-dotted line), and 0.10 (long-dashed line).
	The critical value for the transition from the third family case to a connected hybrid star branch is $\Delta_P\sim 0.06$. 
	The mass ranges ($m_1$ and $m_2$) for the two neutron stars in the merger event GW170817 \cite{TheLIGOScientific:2017qsa} indicate that the heavier could be a third family member while the lighter one is a normal neutron star.}
	\label{fig:M-R-low}
\end{figure}

From GW170817 a constraint for the mass ranges of the merging neutron stars has been derived for the low- and high-spin cases. 
We show the low-spin case in Fig.~\ref{fig:M-R-low} and compare with our class of hybrid star EoSs for $\alpha=0.3$. 
In this case, at least one of the two neutron stars in the merger could belong to the third family branch of rather compact stars. 
This is an exciting possibility which actually has been overlooked in a recent analysis \cite{Annala:2017llu} using a ``generic'' multi-polytrope form of the high-density EoS.
The fact that multi-polytropes can also capture the mass twin case and at the same time fulfill the known constraints for masses and radii of compact stars was recently demonstrated in \cite{Alvarez-Castillo:2017qki}.
 
%%%%%%%%%%%%%%%%%%%%%%%%%%%%%%%%%%%%%%%%%%%%%%%%%%%%

\subsection{Comparison with another third family EoS}

\begin{figure}[!htb]
	\includegraphics[width=1.1\columnwidth]{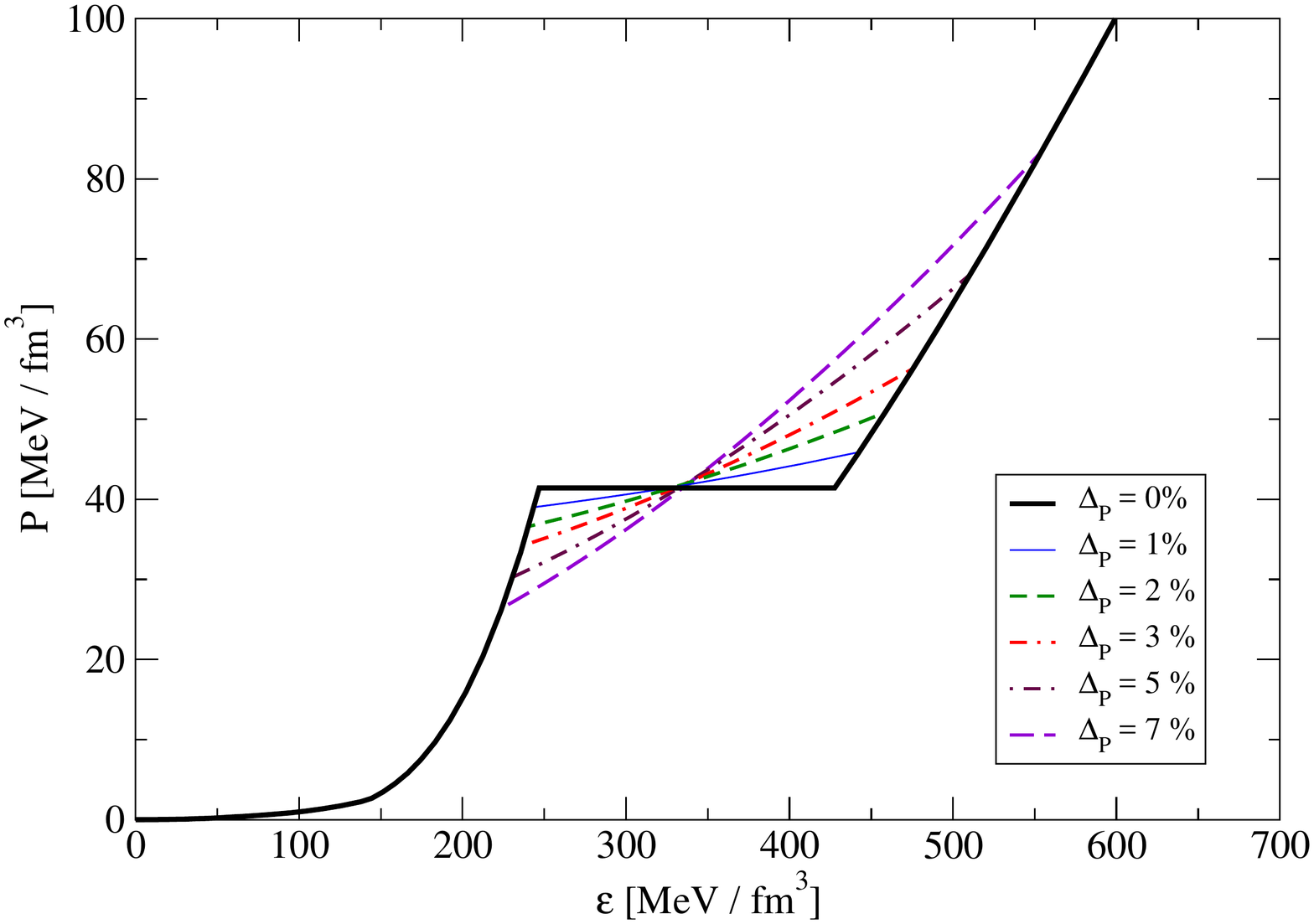}
	\caption{%
	Hybrid star EoS of Benic et al.~\cite{Benic:2014jia} with the mixed phase construction	of the present work applied to it for different values of the mixed phase parameter $\Delta_P=0$ (bold solid line), 0.01 (thin solid line), 0.02 (short-dashed line), 0.03 (double-dash-dotted line), 0.05 (dash-dotted line) and 0.07 (long-dashed line).}
	\label{fig:eos2}
\end{figure}

The question arises whether the quantitative results for the robustness of the third family solutions against the mixed phase construction developed here would also hold when another hybrid star EoS is considered that, under a Maxwell construction for the phase transition, also produces a third family branch in the mass-radius diagram.
In order to answer it, we choose a typical example from the class of hybrid EoSs introduced by Benic et al. \cite{Benic:2014jia} which has a hadronic phase described by the relativistic meanfield model DD2 with a nucleonic excluded volume correction while the quark matter phase is given by a Nambu-Jona-Lasinio (NJL) model with higher order couplings of quark currents that provide the necessary stiffness to stabilize hybrid stars at high mass. 
This EoS, the case $\eta_4=5.0$ from the set discussed in Ref.~\cite{Benic:2014jia}, is shown in Fig.~\ref{fig:eos2} together with results for the mixed phase construction of the present work applied to it for different values of the mixed phase parameter $\Delta_P$. 

\begin{figure}[!htb]
	\includegraphics[width=1.1\columnwidth]{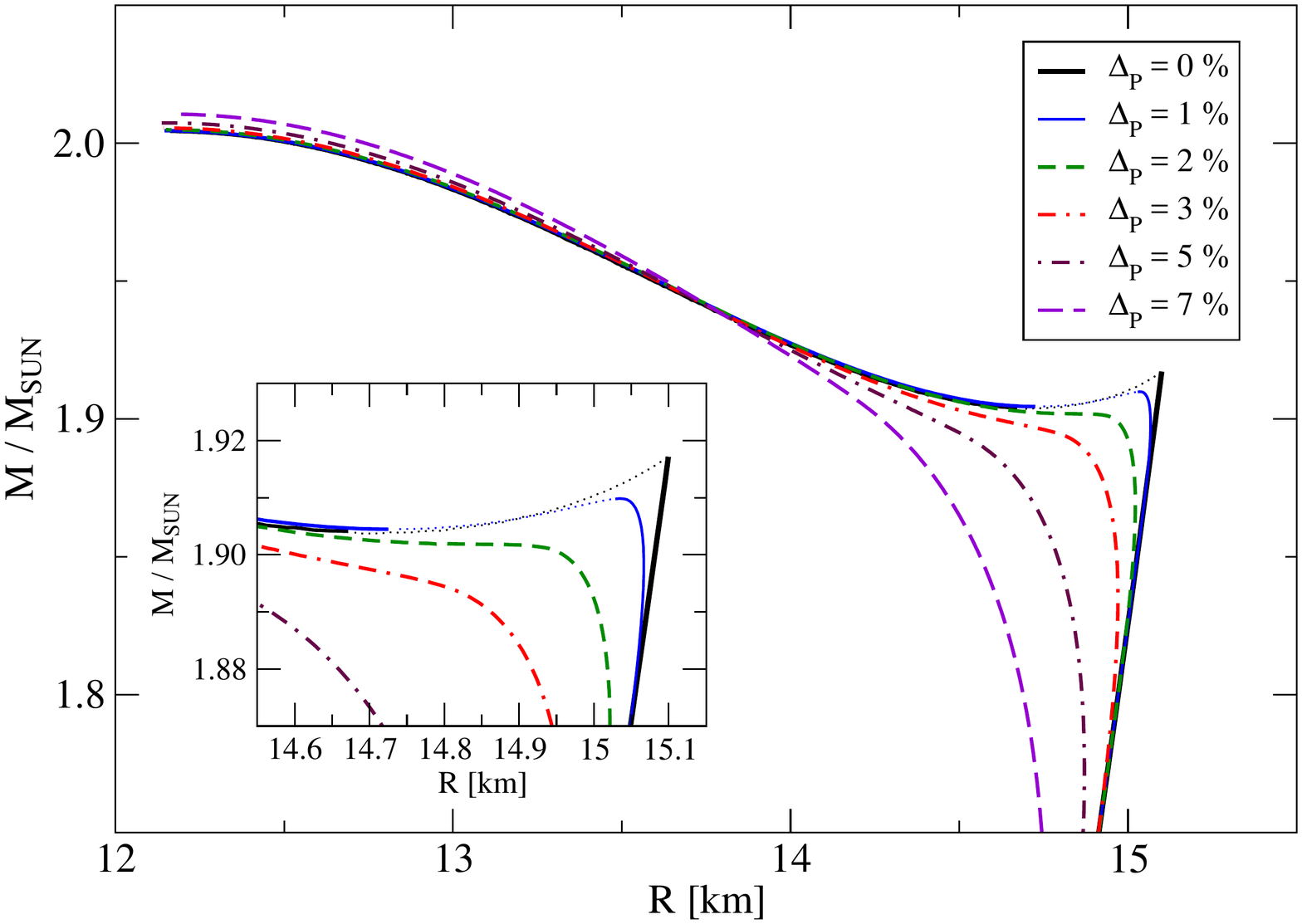}
	\caption{%
	Mass-radius diagram for the hybrid star branches obtained with the EoS shown in Fig.~\ref{fig:eos2} for different values of the mixed phase parameter $\Delta_P$; line styles are the same as in that figure.
	The critical value for the transition from the third family case to a connected hybrid star branch is $\Delta_P\sim 0.02$.}
	\label{fig:M-R-Benic}
\end{figure}

In Fig.~\ref{fig:M-R-Benic} we show the results for the corresponding star sequences in the mass-radius diagram obtained by solving the TOV equations with the EoS for Fig.~\ref{fig:eos2}.   
We observe that already for $\Delta_P=0.02$ the neutron star and hybrid star branches become connected so that the third family and mass twin star solutions exist only for lower values of the mixed phase parameter.
From this exercise it follows that the limiting value of  $\Delta_P=0.06$  for obtaining a third family branch with the new hybrid star EoS of the present paper is not a universal number but is specific to the EoS considered. 
A similar statement holds for the maximal broadening of the phase transition region that would still allow for a third family branch. With the EoS of Benic et al.~\cite{Benic:2014jia}, it amounts to less than half of the critical pressure while for the new EoS of this work it is about the value of the critical pressure itself.

%%%%%%%%%%%%%%%%%%%%%%%%%%%%%%%%%%%%%%%%%%%%%%%
%%%%%%%%%%%%%%%%%%%%%%%%%%%%%%%%%%%%%%%%%%%%%%%

\section{Conclusions}
\label{conclusions}

We have introduced a class of hybrid star EoSs where both, hadronic and quark matter phases are described by a relativistic mean-field theory and would separately satisfy the 2 $M_\odot$ mass constraint for compact stars. 
Performing a Maxwell construction of a first-order phase transition results in a density jump at the transition which is sufficiently large to result in the occurrence of a third family branch of compact stars in the mass-radius diagram.
Varying the screening parameter of the string-flip quark matter model in the range 
$\alpha = 0.18-0.30$ results in a lowering of the onset density for the quark matter phase transition and thus the mass range where pairs of mass twin stars can be found, from $2.05~M_\odot$  to $1.48~M_\odot$, with a reduction of the radius difference from  $3.2$ to $1.5~\mathrm{km}$, respectively.

We have investigated the robustness of the mass-twin (i.e., third family) feature against effects of finite size structures in the mixed phase, so-called ``pasta phases''. 
To this end a phase transition construction mimicking the pasta phase was employed which introduces an additional parameter, the dimensionless pressure shift $\Delta_P$ relative to the critical pressure $P_c$ at the critical chemical potential $\mu_c$ of the Maxwell transition.
We have found that values of $\Delta_P$ up to $5\%$ do not exclude the existence of the third family of compact stars, which was obtained via the Maxwell construction. 
This limiting value corresponds to a lowering of the onset pressure by about $40\%$ and a broadening of the phase transition region in the pressure of the order of the critical pressure $P_c$ for all parametrizations within the present class of models.
These results are specific to the new class of EoSs introduced in this work.
For comparison, we applied the mixed phase construction also to a member of the class of hybrid EoSs that have been introduced by Benic et al.~\cite{Benic:2014jia} and found that in this case the third family solution ceases to exist for $\Delta_P\gsim 0.02$ and the corresponding limiting value for the broadening of the phase transition region amounts to less than half of the critical pressure.
Therefore, we conclude that the class of hybrid EoSs introduced in this work is especially suitable for investigating the case of third family branches in compact star phenomenology. 
The same may hold for the class of hybrid star EoSs introduced by Kaltenborn et al. \cite{Kaltenborn:2017hus} which is based on the same quark matter EoS from the class of relativistic density functional models.

The NJL model based quark matter EoS of ~\cite{Benic:2014jia} requires a very stiff hadronic EoS to obtain third family solutions thus entailing large radii on the neutron star branch.
When the higher order vector coupling is decreased the onset mass for hybrid star solutions gets lowered and third family solutions cease to exist before they could reach the mass range of binary radio pulsars around $\sim 1.4~M_\odot$, as shown in \cite{Benic:2014jia}.   
Consequently, it appears unlikely that such hybrid star EoS can be relevant for a discussion of the binary neutron star merger GW170817 with the mass range of $1.16\le M/m_\odot \le 1.6$.

The situation with the present hybrid star EoS (and also that of Ref.~\cite{Kaltenborn:2017hus}) is quite different as can be seen in Fig.~\ref*{fig:M-R-low}.
At least the heavier of the NS of the binary merger GW170817 could have been a member of the  third family of hybrid stars. 
After the completion of this work, such a scenario was considered in Ref.~\cite{Paschalidis:2017qmb} where it was demonstrated that the LIGO constraints on tidal deformability can be fulfilled by a hybrid star - NS inspiral when the hadronic EoS alone would be too stiff to fulfill this constraint for a NS-NS merger.   

\subsection*{Acknowledgments}
We would like to thank David Alvarez-Castillo, Michal Bejger, Sanjin Benic, Stefan Typel, and Nobutoshi Yasutake for valuable discussions.
N.-U.B. acknowledges support from the Polish NCN under Grant No. UMO-2014/13/B/ST9/02621 for the work contributed in Sec. \ref{ssec:qp} and the Bogoliubov-Infeld Program for collaboration between Polish Institutes and JINR Dubna.
The work of A.A., D.B., H.G., K.M. and D.N.V. has been supported by the Russian Science Foundation under Grant No. 17-12-01427. 

%%%%%%%%%%%%%%%%%%%%%%%%%%%%%%%%%%%%%%%%%%%%%
%%%%%%%%%%%%%%%%%%%%%%%%%%%%%%%%%%%%%%%%%%%%%
%\newpage

\end{document}